\documentstyle[epsfig,12pt]{article}
\begin{document}

%\vspace*{1.0cm}
%\hspace*{10.8cm}VECC-NEX-98002

%\bigskip
%\bigskip

\begin{center}

{\large {\bf A Honeycomb  Proportional Counter
              for Photon Multiplicity Measurement 
           in the ALICE Experiment}} 

\bigskip
\end{center}

\rm
%\medskip
%\begin{frontmatter}
%\vspace{0.5 in}

\noindent{M.M.Aggarwal$^1$,
S.K.~Badyal$^2$,
%A.~Bharti$^5$, 
%A.~Bhasin$^3$,
V.S.~Bhatia$^1$, 
S.~Chattopadhyay$^3$, 
A.K.~Dubey$^4$,\\ 
M.R.~Dutta~Majumdar$^3$, 
M.S.~Ganti$^3$, 
P.~Ghosh$^{3*}$, 
%A.~Gupta$^3$,
%S.K.~Gupta$^5$, 
%V.K.~Gupta$^3$,
A.~Kumar$^1$, 
T.K.~Nayak$^3$, 
S.~Mahajan$^2$,
D.P.~Mahapatra$^4$, 
L.K.~Mangotra$^2$,
B.~Mohanty$^4$, 
S.~Pal$^3$, 
S.C.~Phatak$^4$, 
B.V.K.S.~Potukuchi$^2$, 
R.~Raniwala$^5$, 
S.~Raniwala$^5$, 
N.K.~Rao$^2$,
%S.S.~Sambyal$^3$,
R.N.~Singaraju$^3$, 
Bikash~Sinha$^3$, 
M.D.~Trivedi$^3$, 
R.J.~Veenhof$^6$, 
Y.P.~Viyogi$^{3\#}$}

\medskip

%\vskip 0.2 true cm

\noindent{{$^1$Physics Department, Panjab University, Chandigarh 160014, India};~
{$^2$Physics Department, Jammu University, Jammu 180001, India};~
{$^3$Variable Energy Cyclotron Centre, Calcutta 700064, India };~
{$^4$Institute of Physics, Bhubaneswar 751005, India};~
{$^5$Physics Department,  Rajasthan University, Jaipur 302004, India};~
{$^6$CERN, Geneva, Switzerland; }
{$^*$CSIR research fellow.}}

%\date{\today}

\begin{abstract}

A honeycomb detector 
consisting of  a matrix of 96 
closely packed hexagonal cells, each working as a proportional counter
with a wire readout, was fabricated and 
tested at the CERN PS. The cell depth and the radial dimensions of the cell
 were small, in the range of 
5-10 mm. The appropriate cell design was arrived
at using GARFIELD simulations. Two geometries are described illustrating
the effect of field shaping. The charged particle detection
efficiency and the preshower characteristics have been studied using pion and 
electron beams.
Average charged particle detection efficiency
 was found to be 98~\%, 
which is almost uniform within the
cell volume and also within the array. The preshower data show that 
the transverse size of the shower is in close agreement with the
 results of
simulations for a range of energies and converter thicknesses.

\end{abstract}
----------------------------------------\\
\noindent{$^\#$Corresponding author \\ Variable Energy Cyclotron 
Centre\\ Bidhan Nagar, Calcutta 700064 (India)\\ Tel. +91.33.3370032,
Fax: +91.33.3346871\\e-mail : viyogi@veccal.ernet.in}  

%\end{frontmatter}
\newpage
{\bf Keywords}\\

Proportional counter, preshower detector, ALICE experiment, photon 
multiplicity detector


\newpage
\section{Introduction }

The ALICE experiment is designed for a dedicated study of
heavy ion collisions at the Large Hadron Collider 
 at CERN, in order to study hadronic matter
at high density and temperature and to probe
the deconfinement transition and chiral symmetry restoration \cite{tp}.
The design of the detector systems in ALICE has been based on the assumption 
that charged particle  multiplicities 
 in Pb~+~Pb collisions at a c.m. energy of 5.5~A.TeV could be as high as 
 8000 per unit rapidity at mid-rapidity,
decreasing slowly at larger rapidities.
In such  a high multiplicity environment, a preshower Photon Multiplicity
Detector (PMD) has been designed to measure photon multiplicities
in the forward region and to provide estimates of transverse
electromagnetic energy \cite{pmd_tdr}.  Using these measurements
on an event by  event basis, the PMD will be able to study event shapes
and fluctuations~\cite{wa93_flow,wa98dcc}. 

The basic principle of photon detection using
 the proposed PMD is similar to those of preshower detectors used in 
WA93~\cite{wa93} and WA98~\cite{wa98_pmd} experiments at the CERN SPS.
It consists of a highly segmented detector placed behind a lead
converter of suitable thickness. 
A photon 
 produces an electromagnetic shower on passing through the converter.
These shower particles produce signals in several cells of the
sensitive volume of the detector. Charged hadrons usually affect only one 
cell and produce a signal resembling those of
 minimum ionizing particles (MIPs).
The thickness of the converter is optimized such that the conversion
probability of photons is high and the transverse shower spread is small to
minimize shower overlap in a high multiplicity environment.
A charged particle detector of similar granularity may be placed in front
of the converter to act as veto in order to improve the discrimination between
charged hadrons and photons.

The PMD will use gas as sensitive medium, as other options are either too 
expensive (e.g., scintillator readout) or not compatible with ALICE baseline
detector (e.g., a silicon detector placed close to the vertex resulted in 
increased background into the Time Projection Chamber). The choice of detector
technology for use in preshower applications is dictated by the
consideration that (a) signal from charged particles is  
confined preferably to  one 
cell, and (b) low energy $\delta$-electrons should be prevented  
from traveling to nearby cells and causing cross-talk among adjacent channels.

        The conventional technology of gas detectors employs a large 
uniform gas volume with cathode pad readout. 
In these detectors a single charged particle invariably affects 
more than one pad owing to the capacitive coupling between them 
or due to $\delta$-electrons traveling at large angles.
This  leads to  spreading of the signals to regions much 
beyond the actual shower region.

        Segmentation of the gas volume with material separation 
thus becomes necessary to reduce the effect of  $\delta$-electrons. 
Some attempts have been made in the past to develop such segmented 
proportional chambers~\cite{balagura,cplear}. However, they do not suit 
our requirements of small gas thickness, field homogeneity and 
material thickness.

We have used a honeycomb cellular geometry for segmentation and field shaping.
%        The  detector comprises of a matrix of 
%hexagonal cells. The cell walls form the cathode, and a wire along 
%the longitudinal axis in each cell 
%acts as the anode. The honeycomb matrix is filled 
%with a suitable gas such that each 
%cell acts like a proportional counter. 
Several prototypes of the PMD have been 
fabricated. These have been
tested at CERN using pion and electron beams at various 
 energies \cite{pmd_tdr,notes}. 
The same design will also be used for the PMD in the STAR experiment at 
RHIC \cite{star_pmd}.
In this article we 
present an account of the design, fabrication and the performance 
of these prototypes. 
Section 2 describes the detector design and the cell modeling.
In section 3 we describe the fabrication procedure of the 
honeycomb prototype. The test data are discussed in detail in 
section 4. A summary is presented in section 5.

\section{Design considerations and cell modeling }

The most important requirement for the present detector, apart from high
efficiency for charged particle detection, is that the signals
from charged hadrons be confined to one cell. This will help to minimize
the occupancy in both preshower and veto applications. In addition, the design
is also influenced by the following considerations
 for  a preshower detector :

\begin{itemize}
\item the active volume of the detector should be thin and very close
to the
converter  so that the transverse spread of the shower  is minimized, 
\item low-energy  $\delta$-electrons should be prevented from traveling to
nearby cells, in order  to minimize crosstalk among
adjacent channels,
\item the technology should be amenable to modular design with a minimum of
dead space at the boundaries and should not require a staggered layout,
\item the detector material (gas) should be insensitive to neutrons. In a
hydrogeneous medium neutrons tend to produce large signals due to recoil
protons, which can mimic a photon signal.
\end{itemize}

        Based on the above considerations a honeycomb chamber design
was adopted. The honeycomb body made up of copper forms the common 
cathode and is kept at large negative potential. 
The anode wire is kept  at ground potential, which
 facilitates easy coupling with front-end 
electronics.
The cells are thus physically isolated from each 
other by  copper walls of suitable thickness to contain 
the $\delta$-electrons. Optimum values of cell depth and copper wall
thickness have been obtained from GEANT simulation \cite{geant} of preshower
development in such a detector.

In a normal proportional chamber the aspect ratio (ratio of cell 
radius and wire length) is generally very small.
In the present case smaller cell depth (around 10~mm) and comparable cell
diameter makes the aspect ratio 
relatively  large. This design makes our detector an 
unconventional one, where the edge effects at the wire ends dominate. 
The end effects have a direct bearing on the efficiency of  
charge particle detection. Hence proper field shaping 
had to be done to minimize these end effects~\cite{pmd_tdr}.   
Appropriate cell configuration has been arrived at by cell modeling and 
prototype tests.

        A schematic diagram of the hexagonal unit cell and its cross
section 
are shown in Fig.~1. 
%Fig~\ref{models}. 
%The cell walls form the
%cathode and the wire along the longitudinal axis ($x$~-axis) is the
%anode. 
The cell volume is sealed by FR4 printed circuit boards (PCBs,
referred to as `lids' in the figure).
% which provide the soldering  
%contact for the wire and the readout electronics. 
%A second layer of thin copper clad FR4 plates is provided, 
%the outer layer of which is grounded for shielding effects. 
The cell simulations were carried out using MAXWELL~\cite{maxwell} and 
GARFIELD~\cite{garfield} software packages.  
The detector cell is modeled in MAXWELL according to the parameters
 given in  Table 1.
%~\ref{parameter1}. 
Because of the six fold
axial symmetry of the cell and a mirror symmetry in the plane 
perpendicular to the tube axis, only 1/12  of the whole cell was
modeled in MAXWELL. The field maps for the entire cell were then 
generated by GARFIELD using these symmetry considerations.

Two geometries of the cathode have been studied as shown in Fig.~1.
%Fig.~\ref{models}. 
For the cell shown in Fig.~1(b),
%in Fig.~\ref{models}(b), 
the cathode resembles an open tubular structure. This is referred to 
as open geometry cell. In Fig.~1(c)
%Fig.~\ref{models}(c) 
the cathode of the cell
 is extended 
close to the anode wire. This 'extended cathode cell' will be described later.

The field configuration within the cell is studied with the help of drift line
plots for the two geometries.
An electron drift-line plot is shown for the open geometry cell
in Fig.~2 for two tracks at different distances from the anode 
wire. For tracks a little far away from the wire,
a large fraction of these electrons do not reach the wire and  escape 
the drift medium, as shown by dashed lines in the figure.
The loss in the collection of the primary electrons increases as the
distance from the anode wire increases. This affects 
the efficiency of charged particle
detection. The problem becomes acute in the present case where smaller gas 
depth limits the primary ionization. 

The charge collection on the anode can be improved with proper
field shaping. Guard rings have been traditionally used
for field shaping in gas counters. However fabrication and optimization
of guard rings becomes difficult for large arrays with segmented cells.
As a simple alternative, the cathode was  
extended  on the inner copper surface of the lids (using double sided PCBs)  
up to 2~mm from the wire. 
This is shown schematically in Fig.~1(c). 
%Fig.~\ref{models}~(c). 
This cell is referred 
to as extended cathode cell. 
The thickness of the extended 
portion of the cathode was taken as 50 $\mu$m for GARFIELD simulations. 

The electron drift-line plot for the extended cathode cell is
shown in Fig.~3. It 
shows that the 
electrons generated almost anywhere inside the volume reach the anode. 
Only for a small portion of the  region on  either ends, 
some  electrons still escape the drift medium 
 and  hit the lids,
 as shown by dashed lines. But this region 
amounts to less than about 10~\% of the entire volume. 
This suggests  that the extended cathode design should have better uniformity 
of response within a cell.

\section{ Prototype fabrication }

The construction of the honeycomb prototype involved 
two main steps -- fabrication of the honeycomb matrix and assembly of 
the chamber.

        Copper strips of thickness 0.2~mm 
with small notches  cut  at
regular intervals on both edges were  shaped in the 
form of a half-honeycomb using a precision-machined hexagonal jig. The width
of the strip, varying from 8~mm to 12~mm, determined the gas thickness of the
detector. 
Joining two such half-honeycomb strips by spot-soldering resulted in  
a row of hexagonal cells. In this scheme four sides of the hexagon
had a thickness equal to the 
original thickness of the strip and two sides had double the thickness. 
In the final shape notches appeared on all sides of the hexagonal cells. 
This facilitated smooth gas flow within all cells in the chamber. 
The honeycomb matrix was then cleaned with a soap solution and  
coated with graphite paint to improve the aging properties and to suppress 
after-pulsing. 
A FR4 frame was machined to house the honeycomb. 
Gas flow nozzles were fitted on the sides of the frame.
A photograph of the components of the
chamber is  shown in Fig.~\ref{proto}.

        The honeycomb matrix was sandwiched between 
        two printed circuit boards (PCBs), each of thickness 1.5~mm.
One side of these PCBs had solder islands (1~mm $\times$ 4~mm) at the center
of each cell and thin (0.25~mm) tracks leading to 50-pin  connectors.
While one board contained the full wiring layout of the tracks, the other 
had only solder islands to anchor the anode wires. One PCB was first 
bonded with epoxy to the FR4 frame. The edges of the honeycomb were 
glued to it after checking the alignment. The second 
PCB was then bonded to the other side of the frame, taking care of 
the hole alignment and proper connection of the cathode (honeycomb) 
to the SHV connector.
For the extended cathode prototypes, 
the inner surface of the PCBs consisted of tin-plated copper layer in 
contact with the honeycomb cathode, and annular gaps of insulated rings of 
2~mm radius around the ends of the central anode wire.
%the inner faces of the PCB's consisted of a nickel plated copper 
%surface in contact with the cathode extended towards the anode wire 
%up to a radius of 2~mm.

        Gold plated tungsten wire ( 20 $\mu$m
in diameter), after proper cleaning,
was inserted through the holes of the PCB using a wire insertion 
jig. The jig consisted of a small plastic wire spool on an aluminum frame 
with hypodermic syringe needle fitted to the frame.
The wire was drawn through the capillary of the needle and inserted 
through the PCB holes together with the needle. The needle was 
withdrawn after soldering the wire
on to the rear PCB. The wire was stretched to a tension of 25~g (equivalent
to 30\% of its elastic limit) using a spring loaded slider on the spool,
before being soldered on to the front PCBs. Care was taken while soldering 
to prevent flux creepage into the cell. The holes were then closed by 
high viscosity fast-setting epoxy. Two thin copper-clad printed boards
were again bonded on the soldered surface of the above mentioned PCBs, with
copper on the outer surface to make a proper shielding for the entire 
readout connections. The details can be found in 
Ref~\cite{pmd_tdr}.

Negative high voltage was applied to the honeycomb cathode 
through a high value (2.2 M$\Omega$) non inductive resistor with a high 
voltage decoupling capacitor. 
The readout boards containing front-end electronics 
were connected on the chamber as shown in Fig.~\ref{proto}.

%\subsubsection{Readout: }

\section{ Test setup, analysis and results }

We describe the results for three prototypes of different geometrical 
configurations as given in Table 2.

\subsection{The test beam setup }

        The prototype chamber was tested at the T10 beam-line 
of the CERN Proton Synchrotron.
The response of the detector to  
charged particles was studied using 
7~GeV/c pion beams. The test setup consisted  of three
pairs of cathode strip chambers (CSC) 
with the prototype 
detector placed in between the last two CSCs as shown in the 
Fig.~\ref{test_setup}~(a). The CSCs 
were used for tracking the pion beams.  
Two pairs of crossed scintillators in a 4-fold coincidence 
provided the charged particle trigger. 

The setup for preshower studies is 
schematically shown in Fig.~\ref{test_setup}~(b). A lead converter of 
suitable thickness was kept just in front of the honeycomb chamber. 
A gas based Cerenkov detector 
was used to provide the electron trigger in coincidence with the two 
pairs of the scintillators.

        The prototype chamber was mounted on a stand having horizontal 
and vertical  movements, reproducible to within a millimeter, 
facilitating the study of the response of individual cells in the detector.
The pair of trigger scintillators had an overlap of 
about 1 cm$^2$ which was comparable to the area of each hexagonal cell.
Hence for most of the studies related to the detector response, the chamber 
could be positioned in such a way that the beam profile was confined to 
almost one cell. 
       A steady flow of Ar-CO$_{2}$ gas mixture  was 
maintained in the  prototype honeycomb detector.

The readout of the prototype chamber was carried out using 
GASSIPLEX electronics (1.5 $\mu m$ version) \cite{gass_old}
and a C-RAMS based
data acquisition setup \cite{hmpid_tdr}.
For the prototype detector in the present case, which
generates negative 
pulses, the dynamic range of GASSIPLEX  was 75~fC and the output pulse was 
limited to -1V.
The pedestal was around 160 ADC channels. The  r.m.s deviation 
of the pedestal, which is a measure of the noise in the system, was
 $\sim$1.5 units. 
%Since only one 
%adjustment of threshold is possible for pedestals in one chain of GASSIPLEX 
%chips and all the channels must remain positive, the pedestal was dictated 
%by the channel having the highest  offset. 

\subsection{Tracking and clustering }

        Tracking was done with the help of CSCs. These provided an external
reference for the impact point of the particle on the prototype chamber.
In all cases only single-hit tracks having good reconstruction in all 
the 6 planes were taken. A straight line was fit by least square 
method by taking the centroids of the clusters for every track in the 
3 pairs of CSC's. Accordingly the impact point for every track was projected 
in the  plane of the prototype detector. 
These projected coordinates were then used to form 2~mm
bins to study the variation in the efficiency within the cell.

        Particles entering the chamber can at times affect more than 
one cell. Hence the cell hits were first clustered using a nearest
neighbour clustering algorithm. For every hit cell, two concentric rings 
comprising of 19 cells were scanned to identify contiguous cells which 
might have been affected. The ADC channel contents in all the cells
falling within the cluster were added together to represent the pulse
height corresponding to the total energy deposited by the incident 
particle.
 
\subsection{Characteristics of charged particle detection }

Typical pulse height spectra for two prototypes (having cell depths 
respectively 10 mm and 8 mm)  
are shown in Fig.~\ref{mip}(a) 
along with a fit to
a Landau distribution. 
The detectors were operated at -1520~V and -1450~V respectively.  
The operating gas in both the cases was a
 mixture of Ar and CO$_{2}$ (in the ratio 
70:30 by weight). 
The mean value of the pulse height 
spectra is taken as the measure of the average energy deposited by the 
charged particle. The distribution of the cluster size (number of cells fired)
for charged particles for the above 
operating conditions of the detector is shown in Fig.~\ref{mip}~(b).
The average cluster size is close to unity, suggesting that the energy
deposition is essentially confined to one cell. This satisfies one of the 
basic design criteria of our detector. 

\subsubsection{Optimization of operating conditions }

%The operating conditions of our detector refers to the high negative 
%voltage applied to the cathode and the gas mixture in the detector. 

        The operating conditions for the detector refer to  
appropriate detector bias voltage and the gas mixture. These were  
optimized after a detailed study of variation of the charged particle
detection efficiency in the proportional region. 

Fig~\ref{eff_mean}~(a) shows the variation of efficiency with 
voltage for different gas mixtures using extended cathode prototypes 
of 8~mm gas thickness (prototype-99-8) and 10~mm
prototype (prototype-99-10).
We observe that the efficiency increases with voltage and at
about -1450 Volts, it reaches a plateau region.
The variation of the peak pulse height with the applied   voltage
is shown in  Fig~\ref{eff_mean}(b) for a given gas mixture and cell
geometry. We observe that around the same 
voltage of -1450 Volts the peak pulse height varies almost linearly 
with voltage indicating the onset of the proportional region.

%A mixture of $Ar/CO_{2}$ mixed in the ratio($70/30$) 
%was used as the optimized gas mixture.
%The Fig.~\ref{eff_mean} shows the variation of efficiency to detect 
%charged particles 
%for different operating conditions and the variation of the mean 
%ADC deposited by 7 GeV/c pion at different operating voltages.
%Comparing the 2 plots we find for a high voltage of above -1450V the desired  
%criteria mentioned earlier is satisfied for all the different gas mixtures. 
%Thus the operating voltage was selected to be -1520V. 
A higher percentage of Ar 
in the gas mixture increases the detection efficiency of charged 
particles, but the longer tails in the signal are 
removed faster if the percentage of CO$_{2}$ is higher, which is 
important for handling a higher rate of particle fluxes \cite{sauli}. 
In the present study with a gas mixture of 
Ar and CO$_{2}$ in the ratio $70:30$, stable operation of the detector was 
achieved with  efficiency comparable to those with higher percentage of Argon.
This suggests that the detector may be able to handle high count rates without
loss of detection efficiency. Final optimization of the gas mixture will be
done after a study of the count rate behaviour.

\subsubsection{Variation of efficiency and pulse height within the cell }

A high statistics scan of the chamber was taken for selected cells to study
the variation of efficiency within the cell in smaller bins in horizontal 
and vertical directions.
      Tracking was done using the hits on 
      the CSCs 
      to get the projected hits on the detector plane. These hits 
      were analyzed for sliding square bins of 2~mm width and 2~mm height. 
      Fig.~\ref{x-eff} shows the variation
      of this efficiency with horizontal position (scanned across three cells)
for both open geometry
and extended cathode cells. It shows a flat region 
      in the central part of the cell, indicating the uniformity in
      efficiency. Compared to open geometry cells, the extended cathode cells
give much flatter efficiency as a function of position along the cross-section
of the cell.

      At the edges, while the efficiency drops to 70~\%
      in case of the open geometry cell, it is above 90~\% for 
      the extended cathode geometry. 
      The large drop in efficiency for the open geometry cell at 
      the edges is due to the poor charge collection for  
      tracks at the edges, as observed from the drift line plots 
      for this geometry (see 
      Fig.~2).
      The distance between the 
      consecutive dips  reflects the cell wall separation 
      for both  cell geometries. 
          
          We have also studied the variation of pulse height 
      as a function of the radial distance of the beam 
      tracks. 
      Fig.~\ref{incell_adc} shows the peak 
      pulse height for tracks at varying radial distances from the wire. The 
GARFIELD simulation results, without the effect of electronics, 
are shown by a solid line. The simulation results describe the 
observed behaviour 
reasonably well.
%Slight increase in pulse height for larger $r$ is attributed to
%varying multiplication region near 
%the wire and some losses in the charge collection at small r. 
%The 
%tracks at large r reach the wire in the central region of the anode where the 
%multiplication is higher in comparison to those near the wire in which 
%case some tracks  also end up near the 
%wire ends. 
Considering that the r.m.s. deviation of pedestals was $\sim$1.5 units,
a noise threshold of 6 ADC channels (corresponding to 4$\sigma$ values)
was applied to the pulse height data during analysis.
      The plot clearly shows that the peak pulse height is 
      greater than the noise threshold by a factor of 5 or more, for 
      tracks at all distances from the wire. This suggests that for
the present detector and electronics system, the signal to noise ratio for
minimum ionizing particles is better than 5:1.

%In GARFIELD simulation 
%pulse height is represented by the sum of the avalanches caused by the 
%ionisation electrons generated by a particle track. Solid line in 
%Fig.~\ref{incell_adc} shows the simulation results for 500 pion tracks, 
%each of energy 3 GeV for different track locations. 
%From the values of the peak pulse height we observe that the 
%avalanche size first decreases and then increases as the track 
%position goes from 0.5 mm to 5 mm. This is in accordance with the observation
%that tracks whose electrons drift towards 
%the central region would give higher pulse height as compared to 
%those that end at the wire ends. Thus data and simulation are qualitatively 
% in agreement 

\subsubsection{Cell-to-cell variation of efficiency and average pulse height} 
  
         About 40 cells in the prototype were randomly selected and 
the beam positioned in the center of each cell to study the charged
particle detection efficiency and average pulse height. The relative 
gains of the cells defined by the ratio of the mean pulse height in a
cell to the average value of the mean pulse heights of all the 40 
cells taken together is shown in the Fig~\ref{relative}. The overall 
gain of the prototype chamber was found to be quite uniform, the 
distribution having a narrow width of $\sigma$ $\sim$ $6\%$. The mean 
efficiency of the different cells scanned for the prototype with 
extended cathode and 8~mm cell depth was found to be $\sim 98 \%$ 
as shown in the Fig~\ref{relative}. The average efficiency 
 for the prototype with extended cathode and 
10~mm cell depth was also close to the above value.

\subsection{Preshower characteristics }

The preshower characteristics were studied for the 
extended cathode prototype-99-8 detector using electron beams in the 
energy range of 1--6 GeV at two different operating voltages
using 2~X$_0$ and 3~X$_0$ converter thicknesses. The results
are described below.

%The test beam setup is shown in Fig.~\ref{test_setup}~(b).

\subsubsection{Transverse shower spread }

     The electron beam  passing through the lead converter
produces an electromagnetic shower and thus affects several 
cells in the detector. The average number of cells affected 
gives us an estimate of the transverse shower size.

A typical distribution of the preshower cluster size 
(number of cells affected)  
for 3~GeV 
electrons  with a 3~X$_0$ thick converter   
is shown in Fig.~\ref{cells_3gev}.

The preshower cluster size 
has been studied for
several combinations  of
electron energies and  converter thicknesses.
The results are summarized in Fig.~\ref{preshower_size}.
Comparing the results with the single particle GEANT simulation
for electrons, it was found that
the average preshower cluster 
size is very close to that obtained in simulation.
This is a very important result 
and represents a marked improvement over the WA98 PMD case
where the preshower cluster
size in test data was about twice that obtained in simulation \cite{wa98_pmd}.
The present results suggest that the
final occupancy in the actual experiment will be close to those obtained
in GEANT simulations.

\subsubsection{Energy deposition spectra in a preshower }
       
Total energy deposition in a preshower is represented by the sum of signals
(ADC contents) of all the affected 
cells in a cluster.
Pulse height spectra representing energy deposition 
 for different electron 
energies for a 3~X$_0$  thick converter are shown in 
Fig~\ref{preshower_spec}. The left panels show the test 
data and the right panels show the simulation results.
The two sets of spectra look similar in shape.
The average energy deposition increases with  increasing 
electron energy. 
The relative widths in the preshower spectra are larger than those in 
simulation. This is due to fluctuations 
 in gas ionization, signal generation and transmission 
processes in data, which are not accounted for in GEANT simulation. 
     
\subsubsection{Energy -- ADC calibration relation}

A comparison of test data with simulation allows us 
to understand the features specific to the readout system and 
cross-talks, if any, and to parameterize the effects.          
%The relationship between the average energy depositions in data and
%those from simulations was studied. 
In view of this, the mean pulse height in ADC units as obtained from 
data is plotted against the mean energy
deposition values in keV from the simulation spectra for different electron 
energies and converter thicknesses. Fig.~\ref{calib} shows the calibration 
relation for an operating voltage of -1465~V. 
The response of the detector and readout is seen to be fairly linear in 
the range of energy studied. 
The present study extends to
average energy deposition of $\sim$60~keV, which corresponds to
the energy deposited  by 10~GeV photons in the preshower part of the PMD. 

\section {Summary }
         
        In the present article, we have described the design, 
simulation and the test data of a gas based honeycomb proportional counter for 
photon multiplicity measurements in the ALICE experiment. The design is based
on a copper honeycomb matrix as cathode and anode wires placed at the center
of each cell. The detector is operated using Ar--CO$_2$ gas mixture with
the cathode at a high negative potential, and the 
anode wire at ground potential. The diameter of the cell and the
 depth (length of the wire) are
comparable, being in the range of 8-10~mm. Two geometries of the cell, with
different cathode structures, have been studied.
The  design has been optimized by GARFIELD simulations and by prototype
studies using high energy pion and electron beams.

The cellular design
is found to contain $\delta$-electrons and minimize the spread of the signal
to neighbouring cells. 
The charged particle signal is confined mostly to one cell. This
satisfies one of the basic requirements of our design.
Operating conditions like the cathode voltage and  
the proportion of Ar and CO$_2$ in the
 gas mixture have been optimized by prototype tests. 
The efficiency for open geometry cell drops considerably near the edges of the
hexagon, but for the extended cathode cells,
it is quite uniform throughout the
volume of the cell.
With the extended cathode prototype of 8~mm gas thickness, 
under an optimized condition of gas mixture and operating voltage, 
the average charged particle detection 
efficiency  was found to be $\sim$98\%. 
A study of a large number of cells of the extended cathode variety shows  
that the relative gains and 
detection efficiencies remain almost uniform throughout the detector, 
 cell-to-cell relative gain varying within 6\%.

    The preshower data show that the transverse shower size  is in close 
agreement with single particle GEANT
simulations for a range of converter thicknesses and electron energies.
 Average pulse height of the preshower
follows  a linear relation with energy deposition for a wide range, upto that
expected from 10~GeV photons in the preshower part of the ALICE PMD.

\newpage
\noindent{\bf Acknowledgements}\\

We wish to express our gratitude to the  accelerator crew for 
the excellent performance of the PS at CERN.
We acknowledge the help of 
S. Iranzo and T. Lopez for help in running the MAXWELL code. We are thankful to
H. Gutbrod, 
P. Martinengo,
A. di Mauro, G. Paic, F. Piuz,  
J.C. Santiard,
J. Schukraft and P. Szymanski   
for many useful suggestions and help 
during tests. 
We gratefully acknowledge the financial support of the 
      Department of Atomic Energy, the Department of Science and
      Technology of the
      Government of India
and the CERN PPE Division for this project. One of us (PG) acknowledges the
grant of research fellowship of
 the Council of Scientific and Industrial Research in India.
%\newpage

\newpage

%%%%%%%%  TABLES
\begin{table}
\label{parameter1}
\caption{Parameters of the unit cell for simulation}
\vspace{0.1 in}
\begin{center}
\begin{tabular} {|l|l|} \hline
Cell radius (typical) & 6 mm \\
Cell depth/Wire length (typical) &  8 mm  \\
Radius of anode wire & 10 microns \\
Thickness of inner FR4 plates (the `lids') & 1.5 mm  \\
Thickness of the outer FR4 plates & 1.2 mm \\
Gap between two FR4 plates & 1.2 mm \\
Dielectric constant of FR4  & 4.4 \\
Thickness of cell wall & 0.2 mm \\
Cathode potential & -1400 V \\ \hline
\end{tabular}
\end{center}
\end{table}

\begin{table}
\begin{center}
%\label{config}
\caption{Geometrical parameters of the prototypes tested.}
\vskip 4mm
\begin{tabular} {|c|c|l|l|} \hline
Cell cross-section & Cell depth & Cathode design & Reference in text \\ \hline
130 sq.mm. & 12 mm & Open geometry & prototype-98 \\
100 sq.mm. & 10 mm & Extended cathode & prototype-99-10\\
100 sq.mm. & 8 mm & Extended cathode & prototype-99-8 \\ \hline
\end{tabular}
\end{center}
\end{table}

\newpage

\noindent{\bf Figure Captions}
\begin{enumerate}

\item{(a) Schematic of a hexagonal cell with coordinate representations
  following GARFIELD convention.
 (b) Cross-section of the cell in $xz$-plane
with dimensions used in GARFIELD simulations. The cathode is limited to the
cell wall. This is referred as `open geometry cell'.
 (c) Section showing the modified cathode design,
 extended on the inner face of the lids
and brought close to the anode wire. The cell with this shape is referred as
 `extended cathode cell'.}

\item{Electron drift lines inside the cell volume in 
$xz$-plane  for an  open geometry cell. The thick line in the center 
represents the anode wire. 
Drift lines for a track 3~mm away from the anode   
are shown on the lower part of the figure and those 
for a track 5~mm from the anode are shown in the upper part. 
Dashed lines represent drift lines escaping the drift medium and falling on the
body of the cell. }

\item { Electron drift lines inside the cell volume in
$xz$-plane  for an  extended cathode cell. The thick line in the center 
represented anode wire. 
%Drift lines for a track 3~mm away from the anode   
%are shown on the lower part of the figure and those 
%for a track 5~mm from the anode are shown in the upper part. 
Dashed lines represent drift lines escaping the drift medium and falling on the
body of the cell. }

\item {Components of a prototype honeycomb proportional counter showing the
honeycomb array, the PCBs and the GASSIPLEX board. In the assembled version,
the PCBs form part of a gas-tight chamber having a high voltage
connection and inlet/outlet for the gas. The gas inlet is visible in the 
picture.}

\item {Schematic of the test beam arrangement. (a) setup 
for pion trigger, (b) setup for preshower studies with a lead converter 
(shown as dark block) in
front of the honeycomb detector (PMD) 
and a Cerenkov counter for electron trigger.}

\item{ Response of the extended cathode prototype to a 7~GeV
pion beam. 
Left part :  pulse height spectra  for the two  prototypes
along with
fits using the Landau distribution  shown by solid lines. Operating voltages
are shown in the figure.
Right:  Distribution of cluster size for pions passing through
 prototype-99-8. }

\item{(a) Variation in the efficiency of charged particle  
detection with applied voltage for different prototypes and gas mixtures,
 (b) Variation
of the peak pulse height (ADC) as a function of applied voltage for 
the prototype-99-8.
}

\item{Variation of efficiency with position within a cell 
for three different prototypes of open geometry and extended cathode cells.
The figure shows a scan of three cells encompassing both sides of the 
boundaries of the central cell.}

\item {Variation of the peak pulse height  with radius $r$ for the
extended cathode cells (prototype-99-8)
 along with GARFIELD simulation results.}

\item {(a) Relative gains of various cells in prototype-99-8
extended cathode geometry, and
(b) efficiencies  of the cells.}

\item{Distribution of cluster size for a typical preshower of 3~GeV
electrons passing through a 3~X$_0$ thick lead converter.}

\item{Average cluster size for electron preshower at different 
energies. The operating voltages and the
converter thicknesses used are shown in the figure. 
The filled circles denote test data and the open circles are GEANT
simulation results.}

\item{Preshower spectra for different electron energies for a 3~X$_0$
converter thickness. The left panel shows the test data while the 
right panel shows simulation results.}

\item{Average cluster signal (ADC)
 vs. energy deposition for various combinations 
of electron energy and converter thickness used in 
the preshower studies at 1465 V.}
\end{enumerate}

\clearpage
\newpage

\begin{figure}
\begin{center}
\epsfig{figure=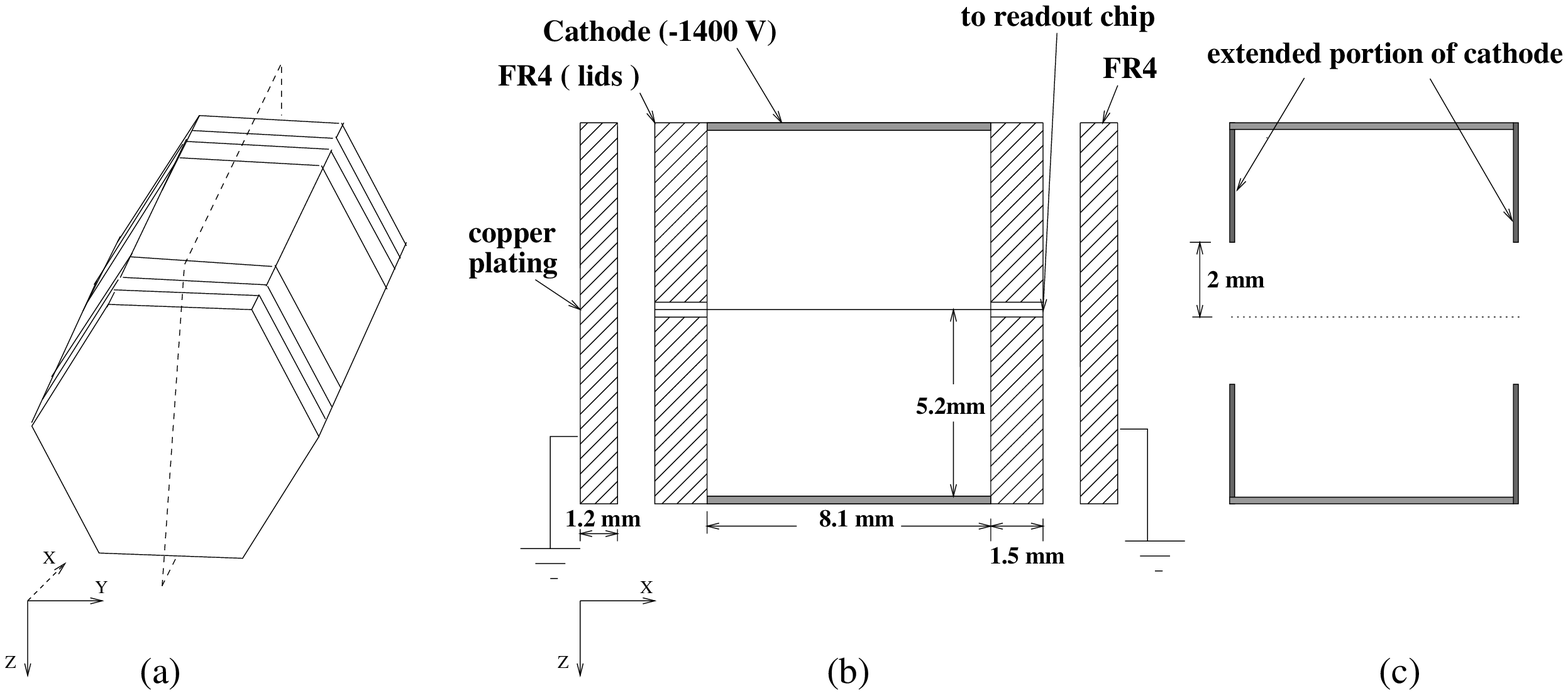,height=6.5cm,width=14.4cm}
\caption{}
\end{center} 
\label{models}
\end{figure}

\begin{figure}
\begin{center}
\epsfig{figure=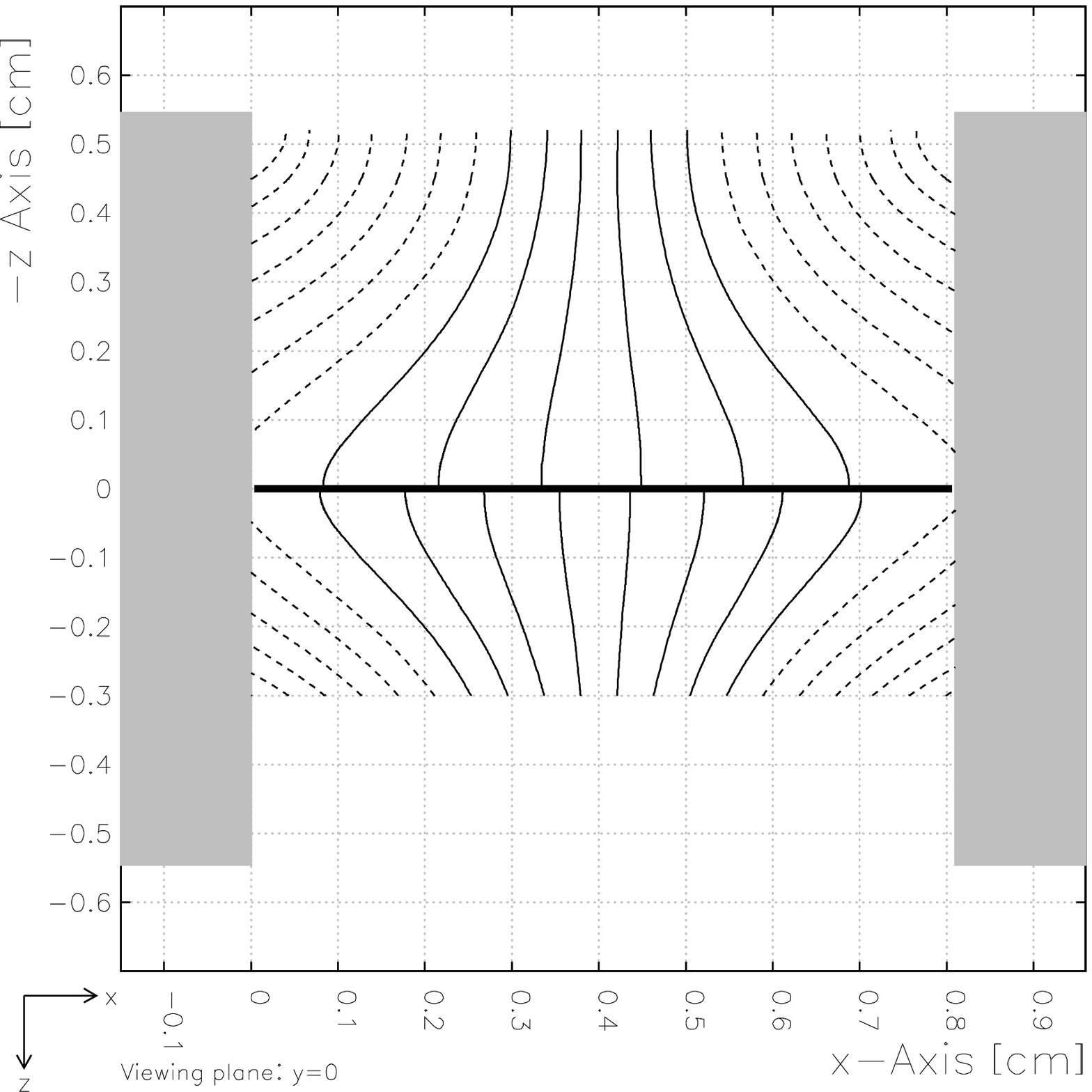,height=8cm,width=8 cm}
\caption{}
\label{drift_o}
\end{center} 
\end{figure}

\begin{figure}
\begin{center}
\epsfig{figure=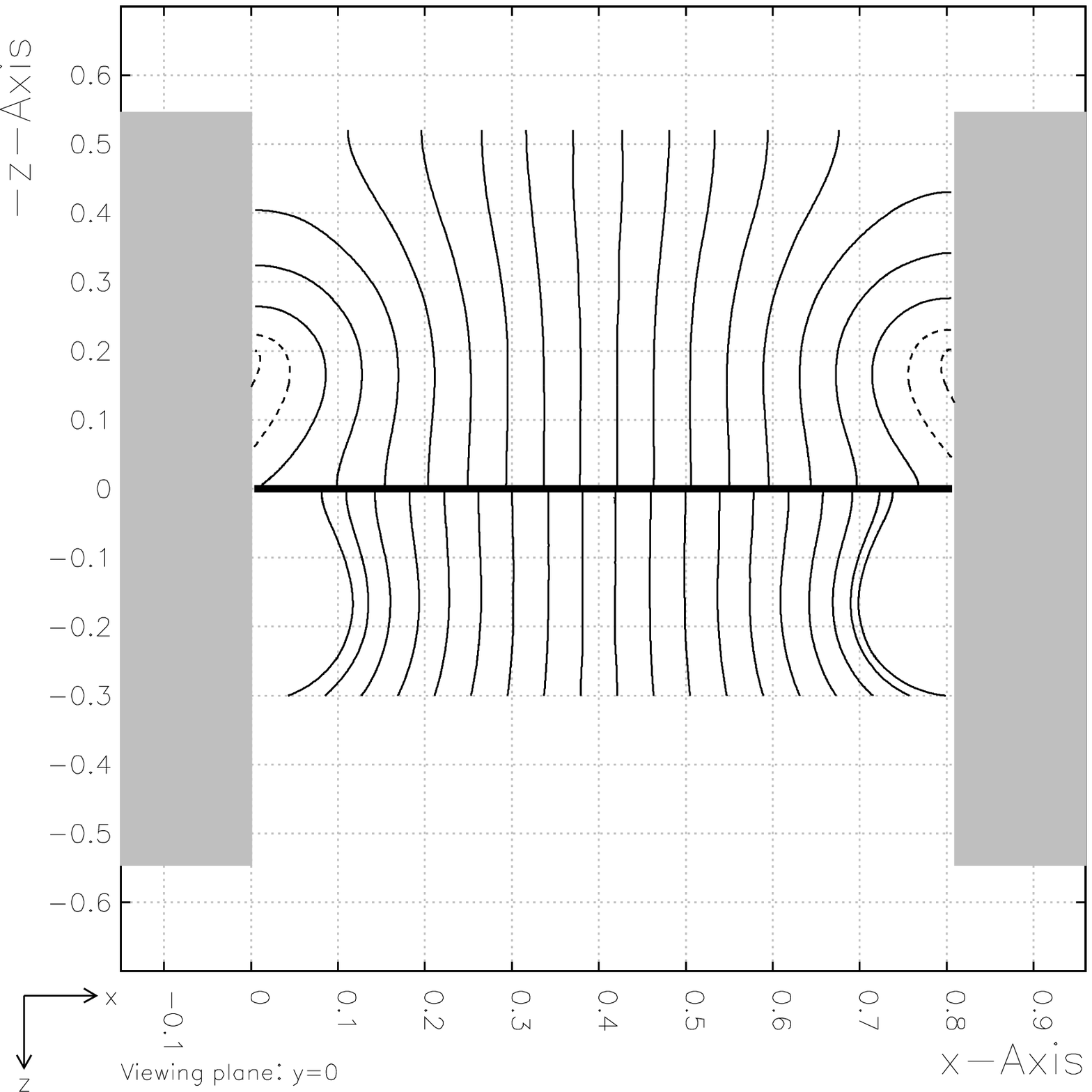,height=8cm,width=8cm}
\caption{}
\label{drift_e} 
\end{center} 
\end{figure}

\begin{figure}
\begin{center}
\epsfig{figure=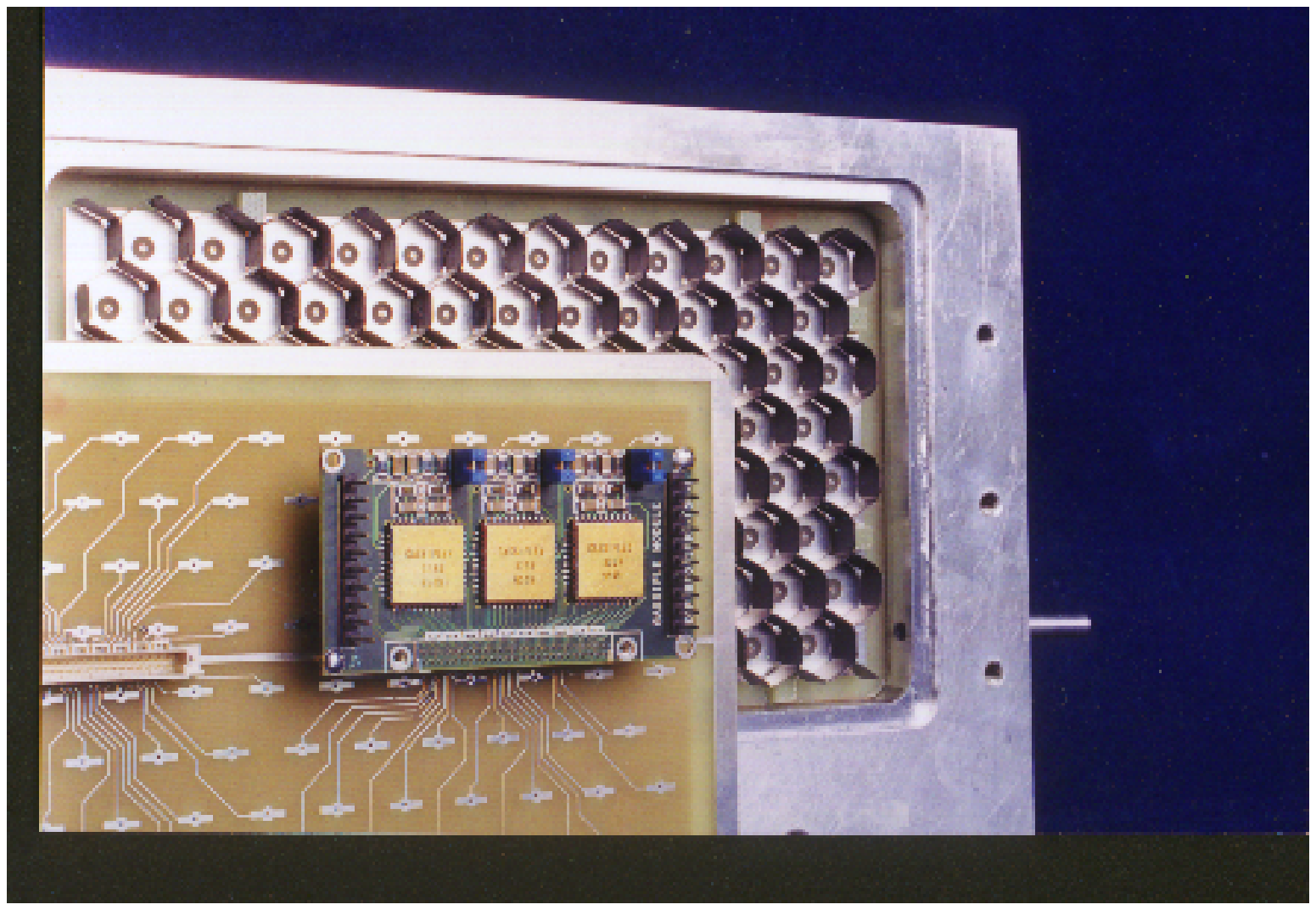,width=14cm}
\caption{}
\label{proto}
\end{center} 
\end{figure}

\begin{figure}
\begin{center}
\epsfig{figure=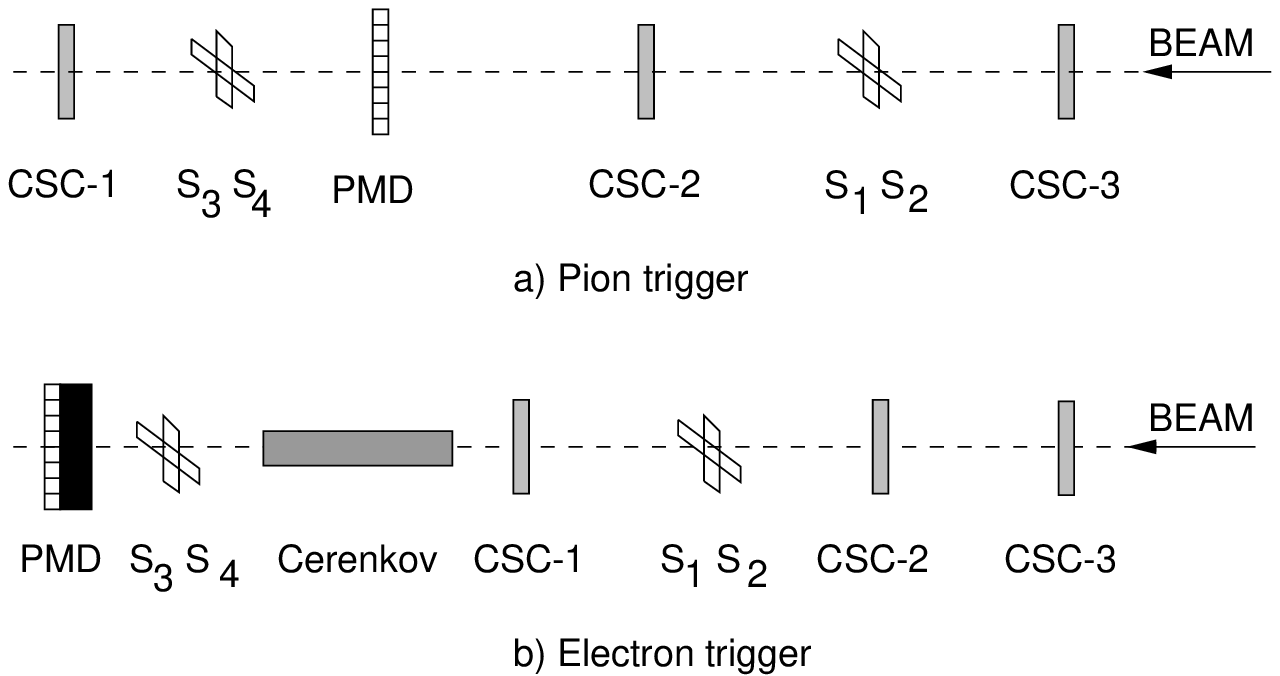,width=14cm}
\caption{}
\label{test_setup}
\end{center} 
\end{figure}

\begin{figure}
\setlength{\unitlength}{1mm}
%\vspace{-0.4cm}
\begin{picture}(80,80)
\put(0,0){
\epsfxsize=8.0cm
\epsfysize=8.0cm
\epsfbox{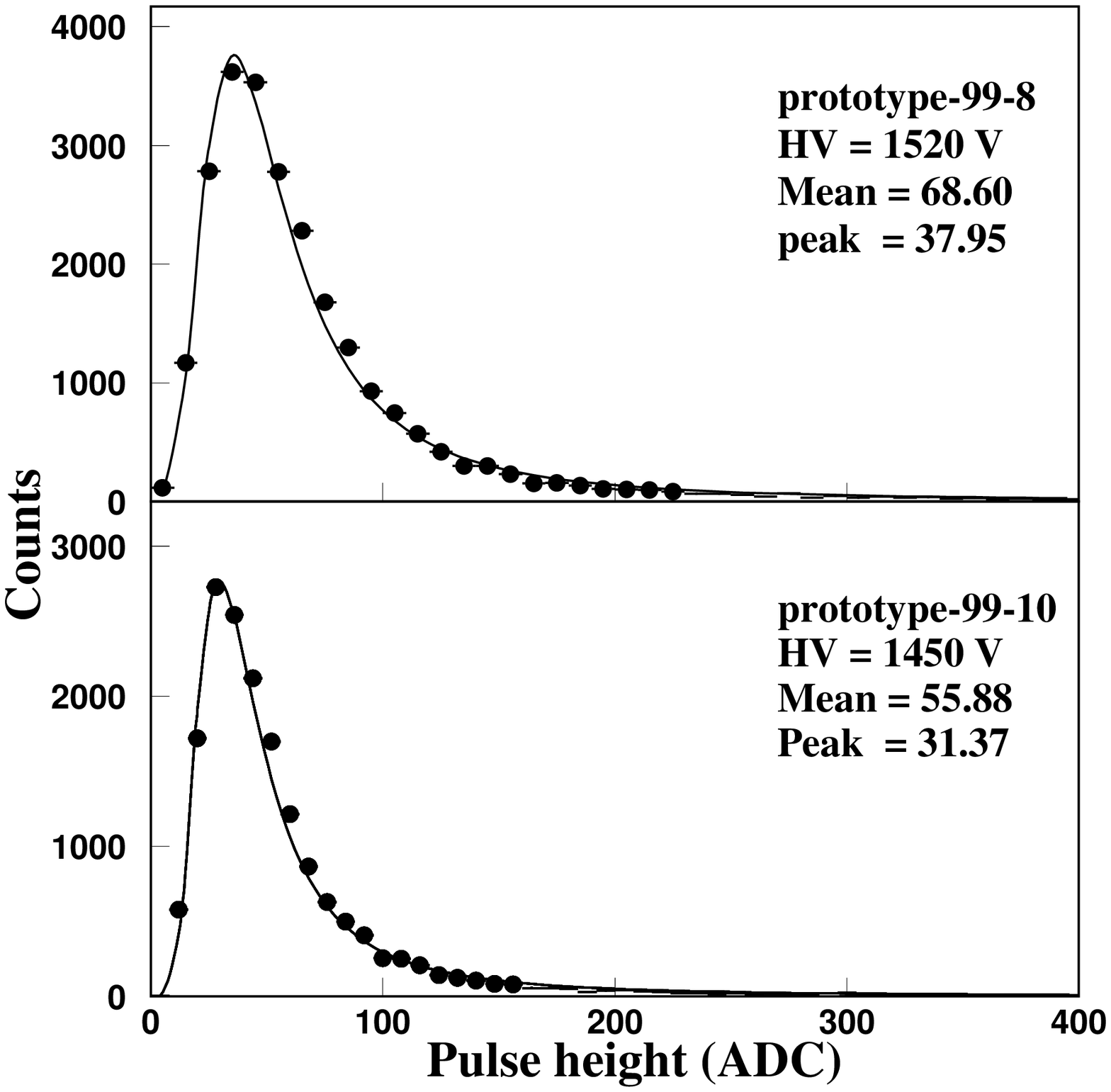}
}
\put(75,0){
\epsfxsize=8.0cm
\epsfysize=8.0cm
\epsfbox{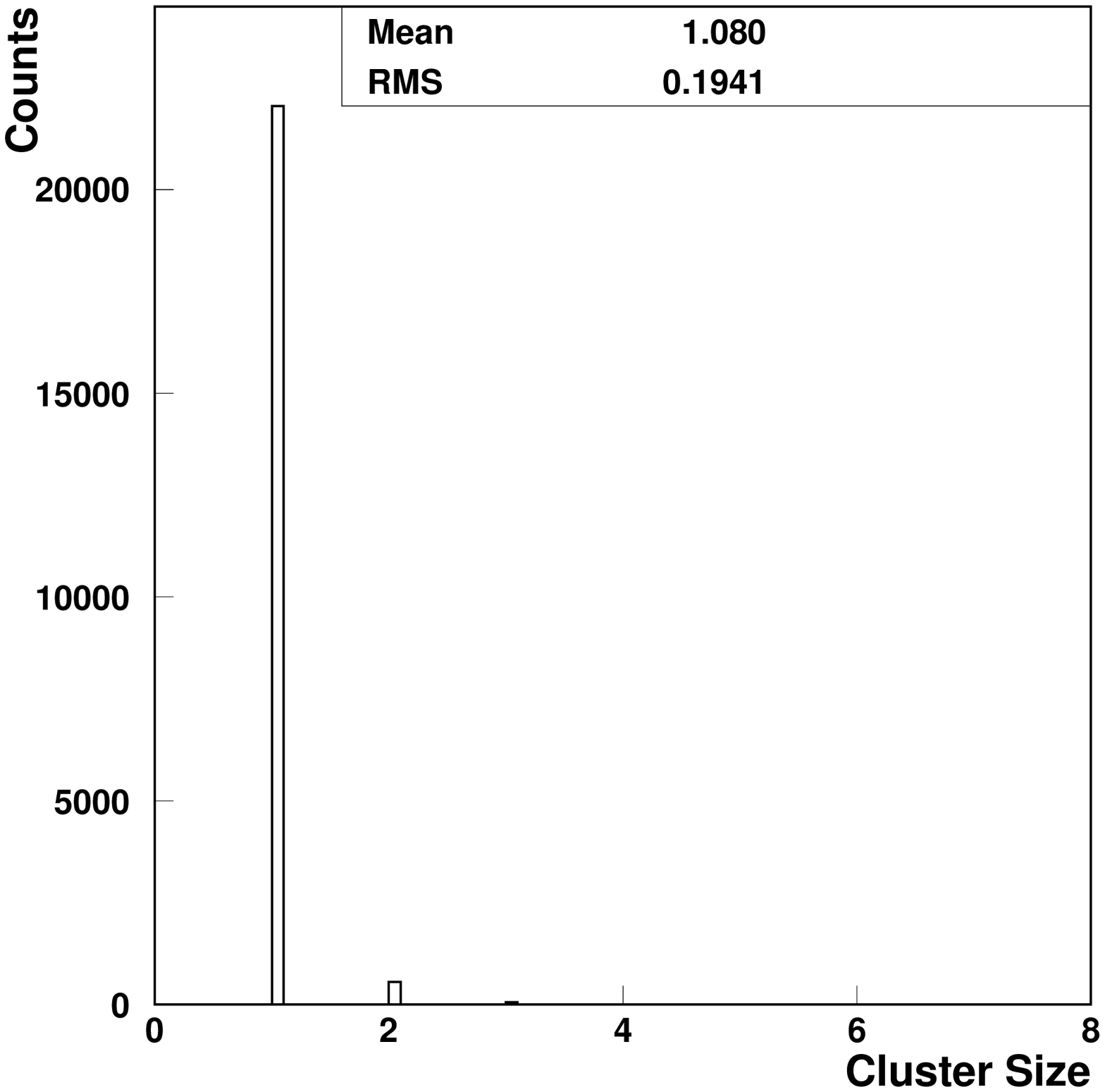}
}
\end{picture}
%\vspace{-1.0cm}
\caption{}
\label{mip}
\end{figure}

\begin{figure}
\setlength{\unitlength}{1mm}
%\vspace{-0.4cm}
\begin{picture}(80,80)
\put(0,0){
\epsfxsize=8.0cm
\epsfysize=8.0cm
\epsfbox{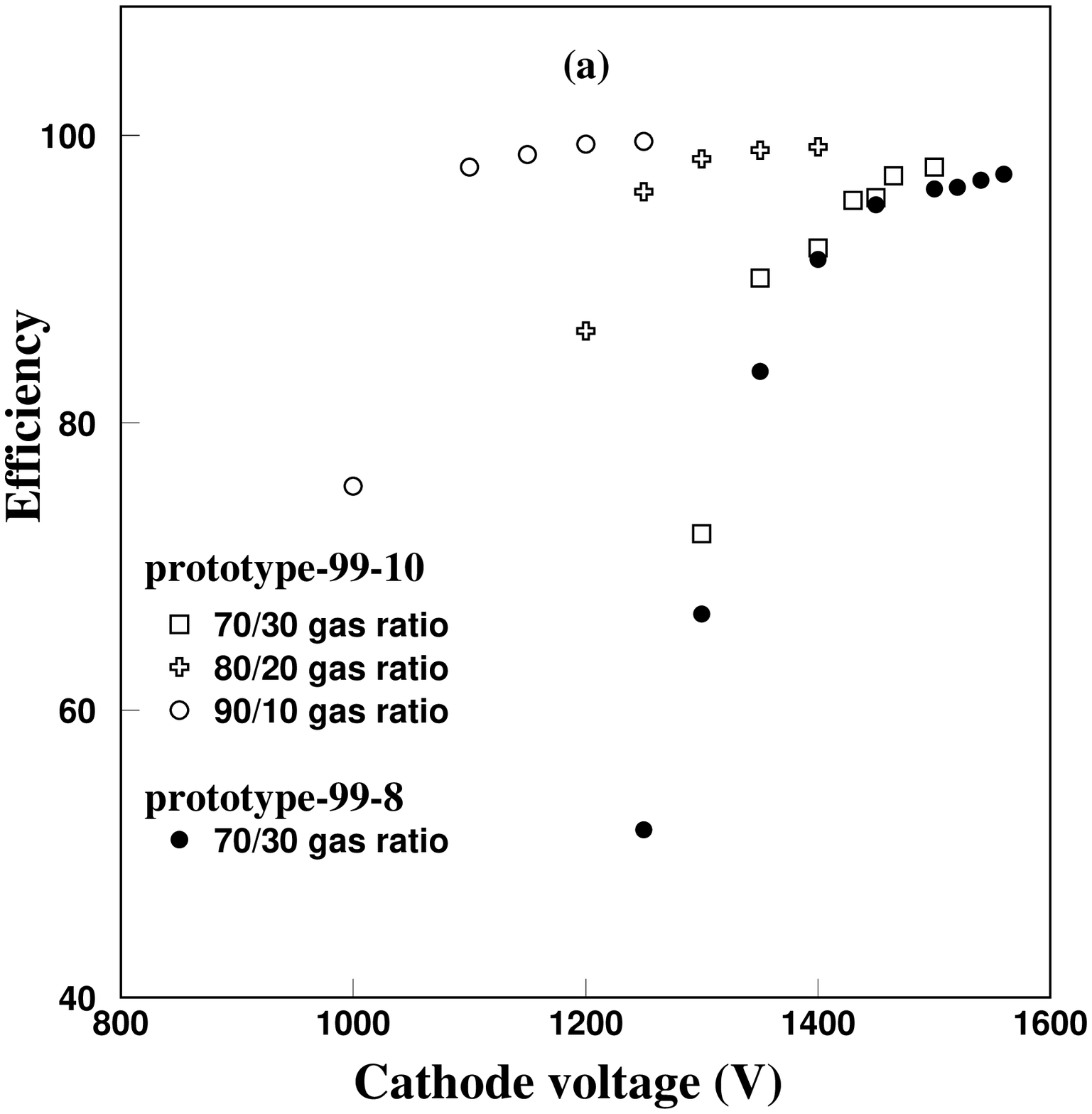}
}
\put(75,0){
\epsfxsize=8.0cm
\epsfbox{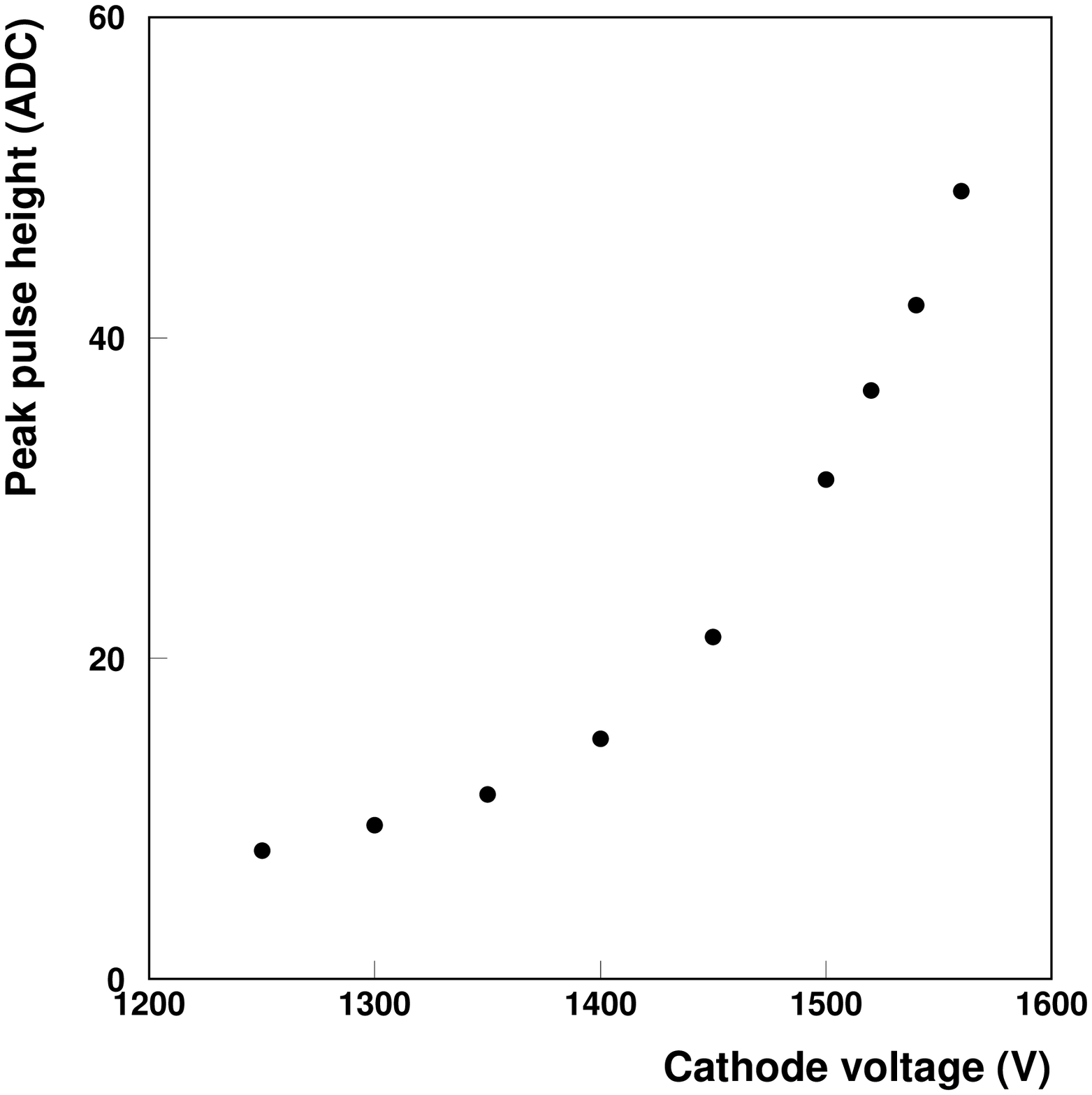}
}
\end{picture}
\caption{}
\label{eff_mean}
\end{figure}

\begin{figure}
\begin{center}
\epsfig{figure=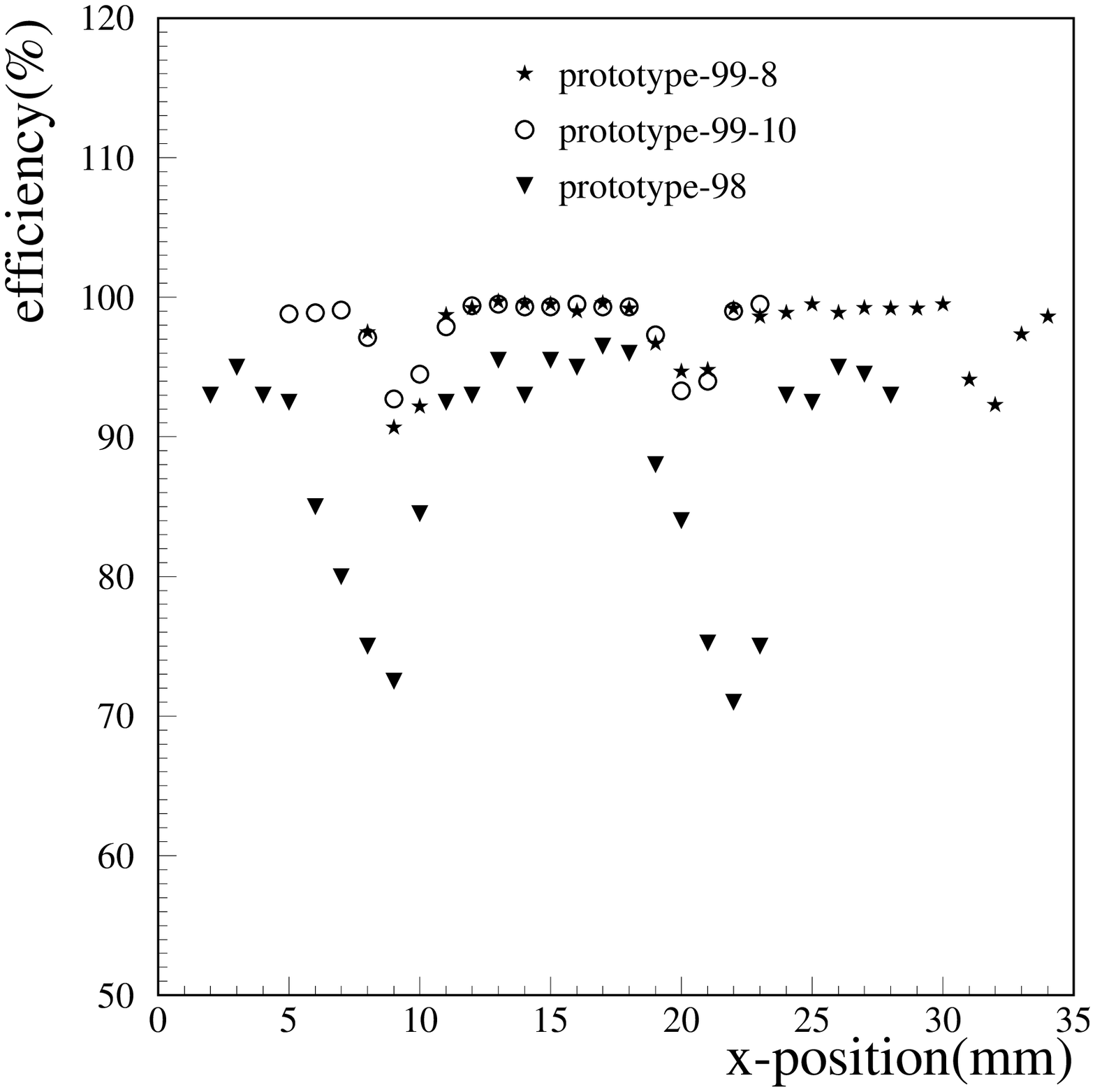,width=10cm}
\caption{}
\label{x-eff}
\end{center} 
\end{figure}

\begin{figure}
\begin{center}
\vspace{3.0cm}
\epsfig{figure=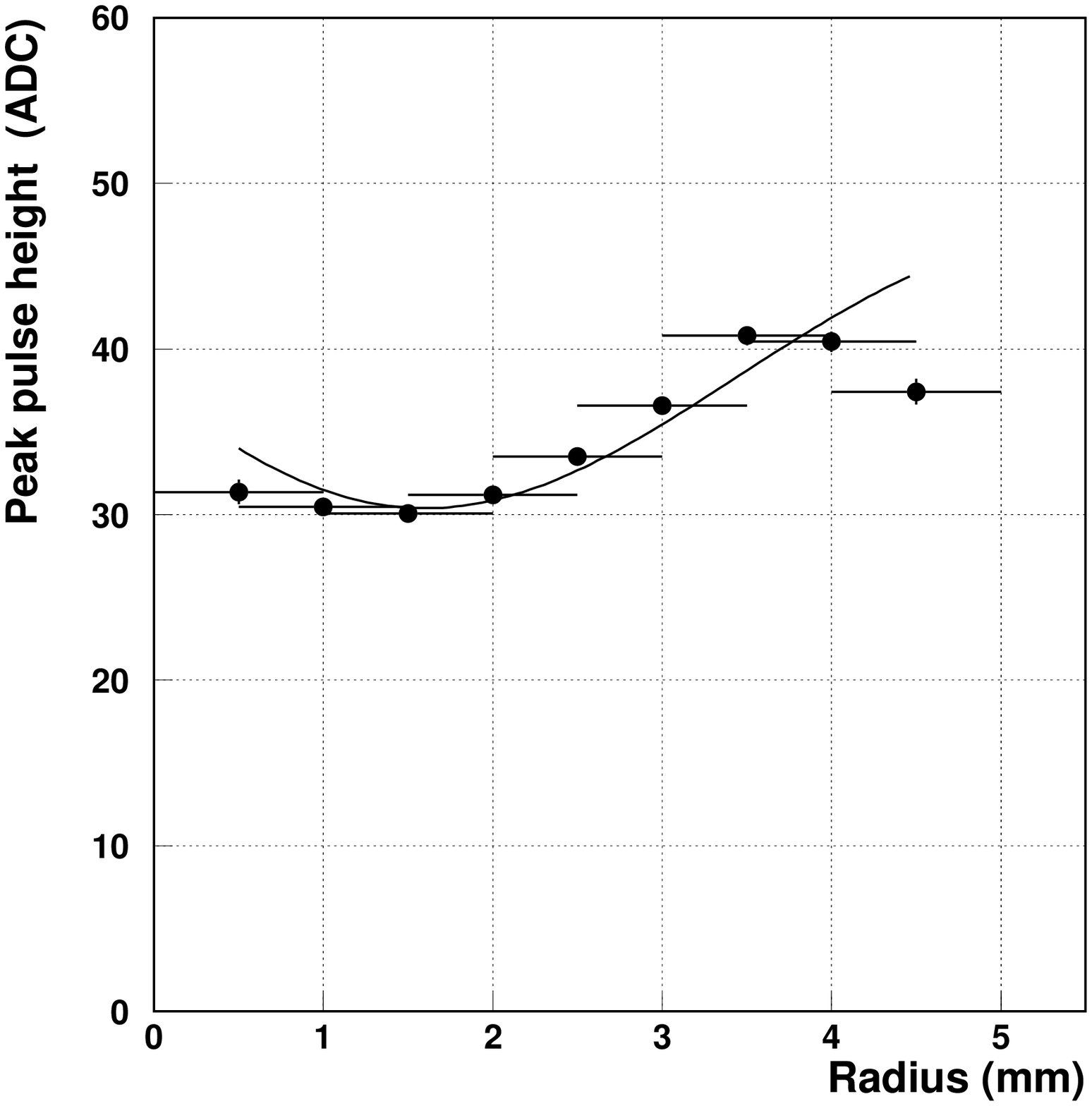,height=10cm,width=10cm}
%\vspace{-2.0cm}
\caption{}
\label{incell_adc}
\end{center}
\end{figure}

\begin{figure}
\begin{center}
\epsfig{figure=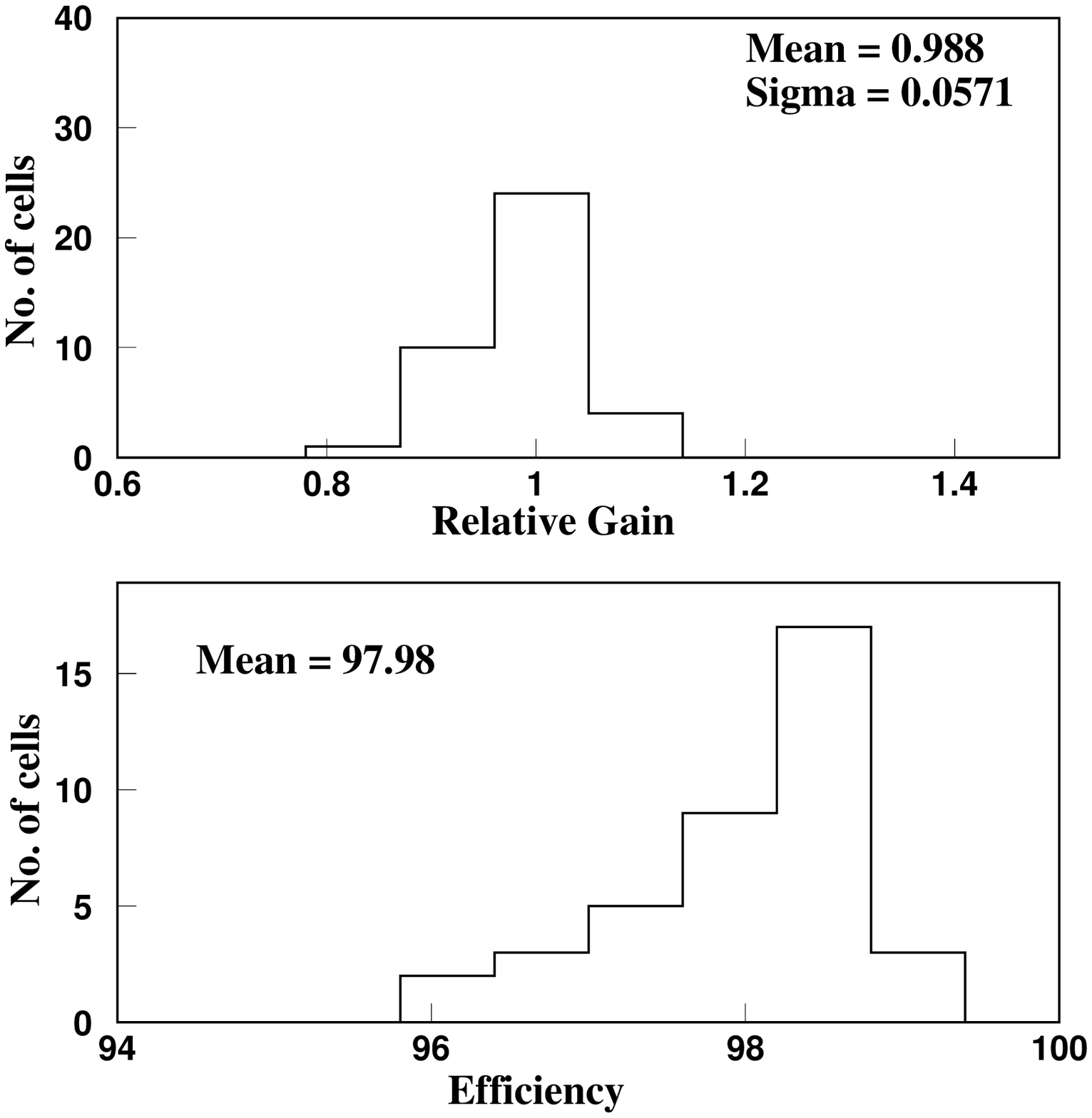,width=14cm}
\caption{}
\label{relative}
\end{center}
\end{figure}

\begin{figure}
\centerline{\epsfig{file=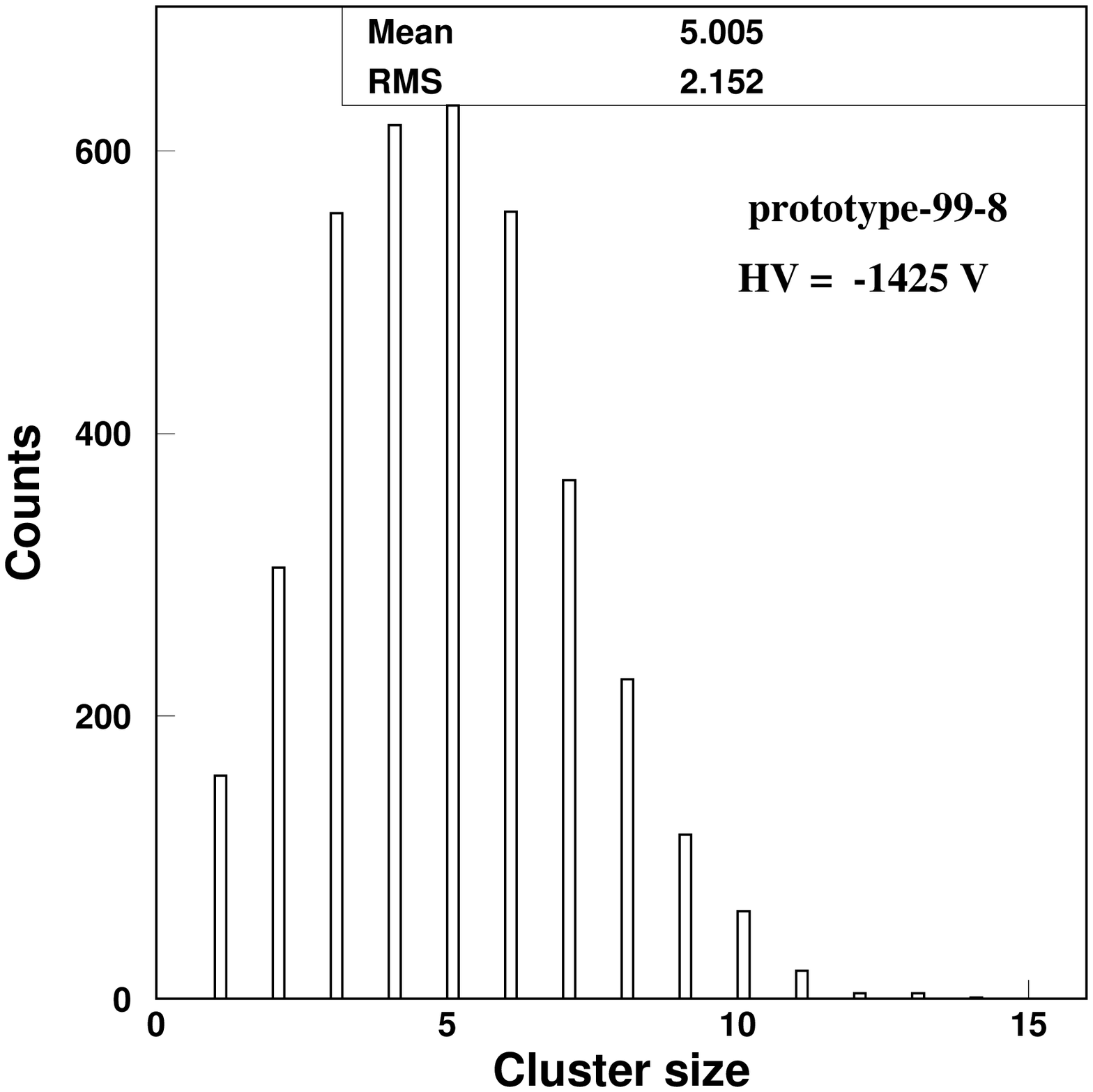,width=8cm}}
\caption{}
\label{cells_3gev}
\end{figure}

\begin{figure}
\centerline{\epsfig{file=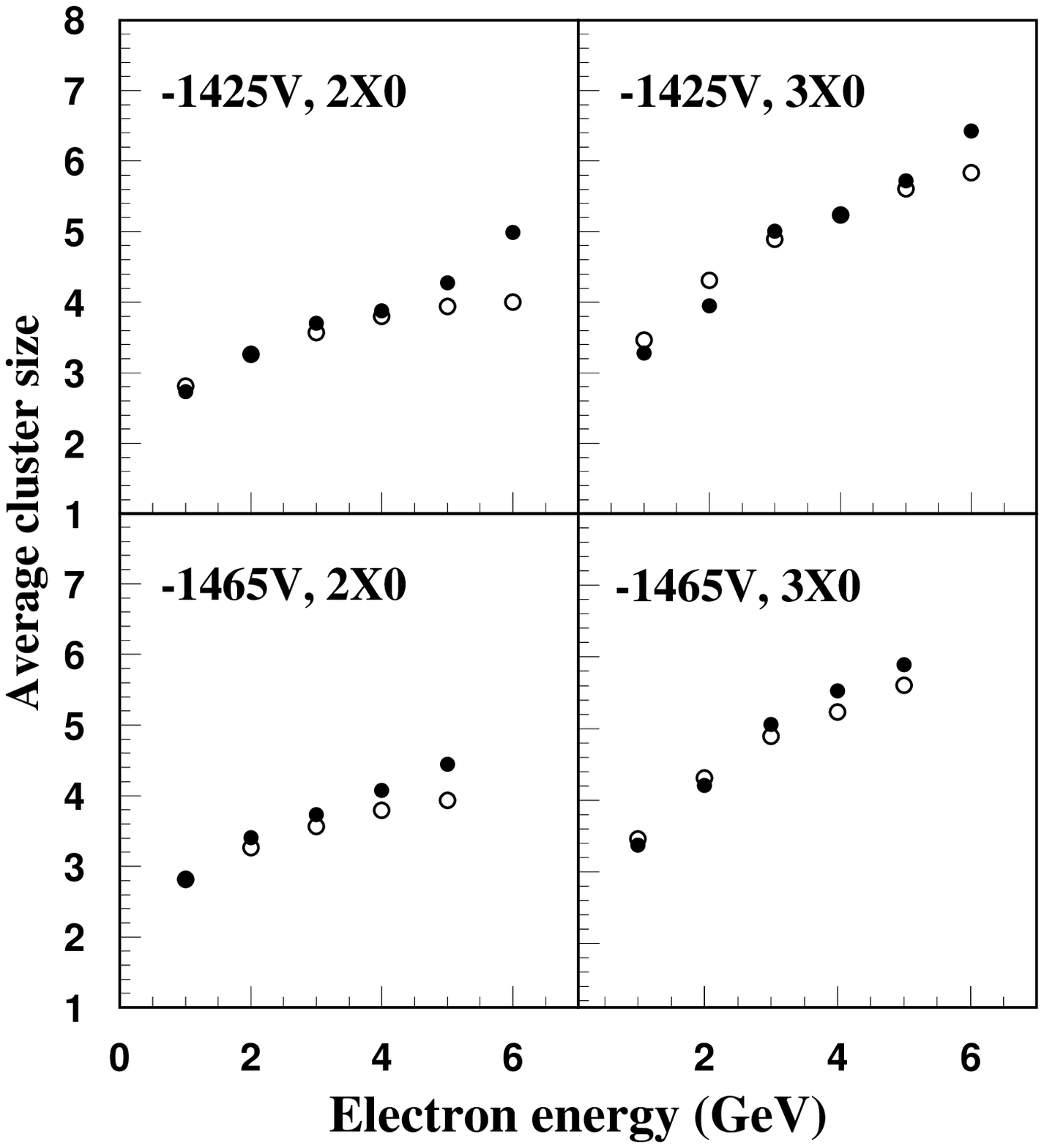,width=10cm}}
\caption{}
\label{preshower_size}
\end{figure}

\begin{figure}
\centerline{\epsfig{file=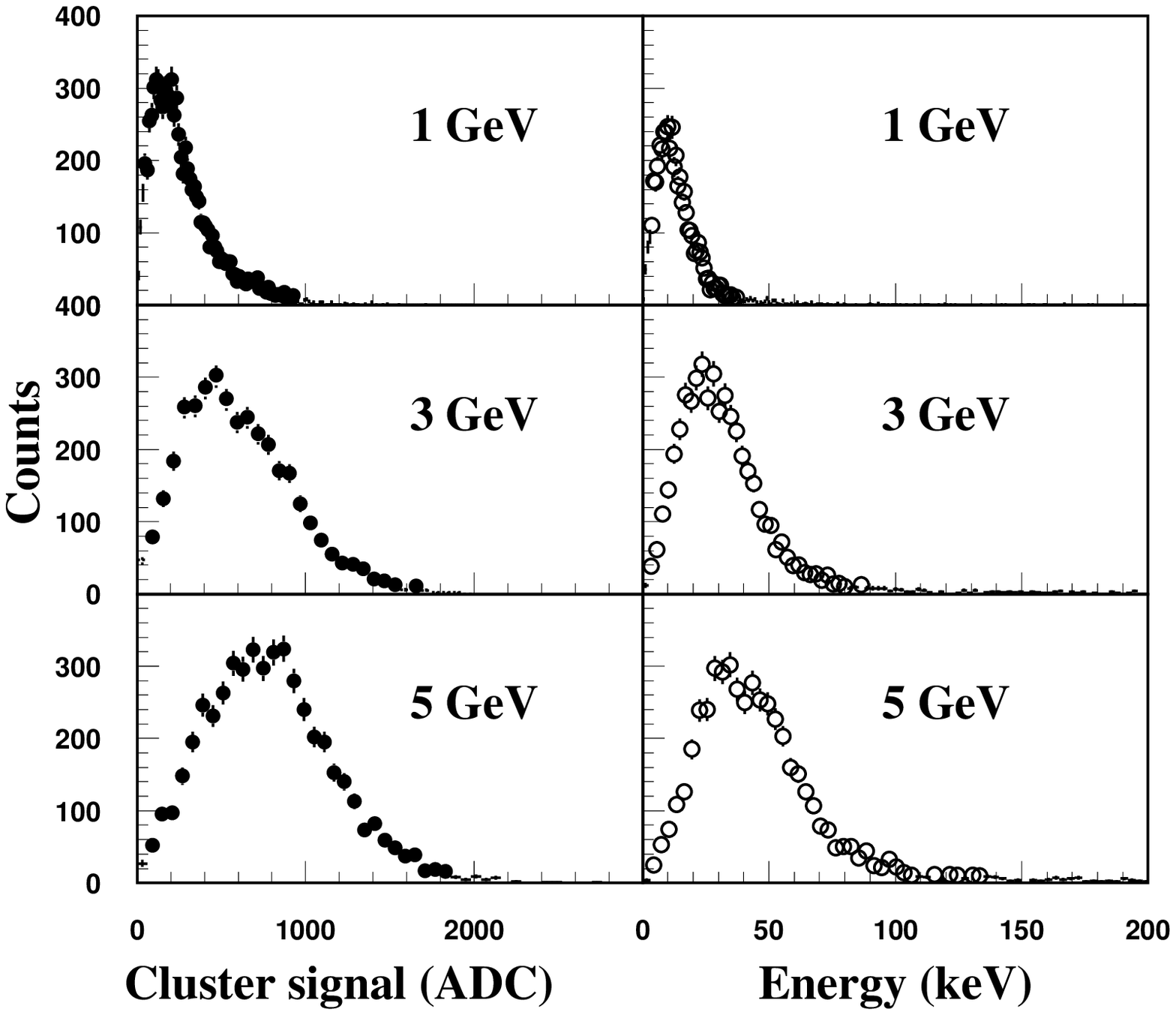,width=12cm}}
\caption{}
\label{preshower_spec}
\end{figure}

\begin{figure}
\centerline{\epsfig{file=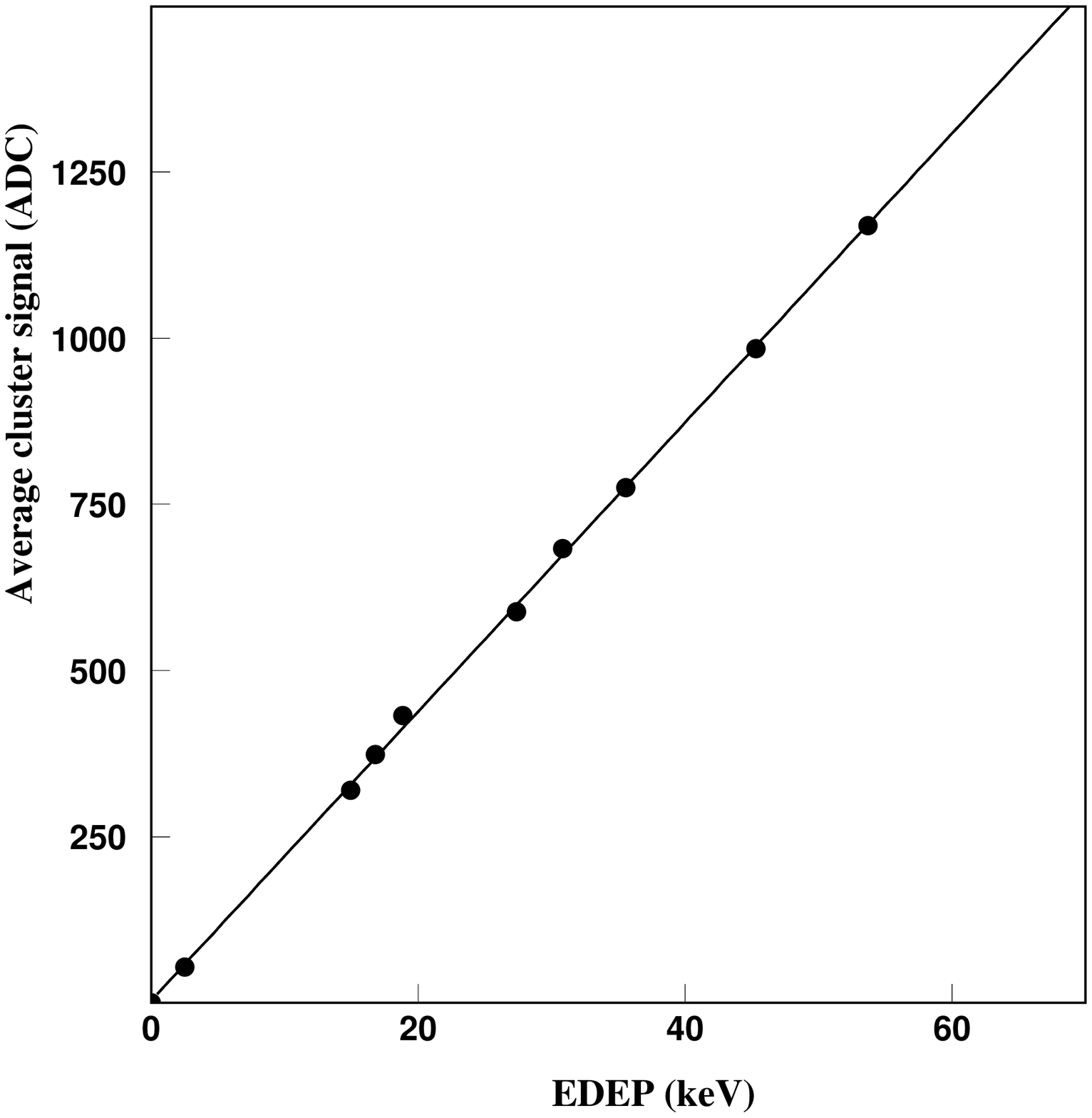,width=12cm}}
\caption{}
\label{calib}
\end{figure}

\end{document}